\documentstyle[aps,twocolumn,10pt]{revtex}
\def\eq#1{(\ref{#1})}
\def\Eq#1{Eq.~(\ref{#1})}

\def\s0#1#2{\mbox{\small{$ \frac{#1}{#2} $}}}
\def\0#1#2{\frac{#1}{#2}}
\def\beq{\begin{equation}}
\def\eeq{\end{equation}}

\begin{document}
\twocolumn[\hsize\textwidth\columnwidth\hsize\csname
@twocolumnfalse\endcsname
\title{Transport theory for a two-flavor color superconductor}
\author{
\hfill 
Daniel F. Litim
and 
Cristina Manuel
\hfill\raisebox{21mm}[0mm][0mm]{\makebox[0mm][r]{ 
CERN-TH-2001-073}}%
}
\address{${}$Theory Division, CERN, CH-1211 Geneva 23, Switzerland.}
 
\maketitle

\begin{abstract}\noindent
{QCD with two light quark flavors at high baryonic density and low
temperature is a color superconductor. The diquark condensate partially
breaks the $SU(3)$ gauge symmetry down to an $SU(2)$ subgroup. We study
thermal fluctuations of the superconductor for temperatures below the gap
$\Delta$. These are described by a simple transport equation, linked to a
quasiparticle behavior of the thermal excitations of the condensate. When
solved in the collisionless limit and close to equilibrium, it gives rise
to the ``hard superconducting loop'' (HSL) effective theory for the
unbroken $SU(2)$ gauge fields with momenta $k\ll \Delta$. This theory
describes Debye screening and Landau damping of the gauge fields in the
presence of the diquark condensate. We also explain how our effective
theory follows to one-loop order from quantum field theory.  Our approach
provides a convenient starting point for the computation of transport
coefficients of the two-flavor color superconductor.\\[1ex]

PACS numbers: 12.38.Mh, 24.85.+p
}
\end{abstract}
\vskip2.pc]

Quantum chromodynamics (QCD) under extreme conditions displays a very rich
phase structure. At low temperature and high baryonic density, quarks form
Cooper pairs due to the existence of attractive interactions amongst
them. The diquark condensate modifies the ground state of QCD and leads to
the phenomenon of color superconductivity
\cite{Barrois:1977xd,Bailin:1984,Alford:1998,Rajagopal:2000wf}. This phase
is typically characterized by the Anderson-Higgs mechanism and an energy
gap associated to the fermionic quasiparticles. It is expected that this
state of matter is realized in compact stars. Studying the properties of
cold dense quark matter may explain some basic features of neutron stars
and could even lead to the prediction of new stellar objects (see
\cite{Rajagopal:2000wf} for a recent review on the subject).

A lot of progress has been made in the microscopic understanding of
color superconductivity. Important open questions concern macroscopic
properties of dense quark matter. Transport theory is known to provide
a very efficient framework for the study of low-energy and long-range
fluctuations in a medium. In this Letter, we present a transport
equation for the gapped quasi-particle excitations of the two-flavor
color superconductor. This is a convenient starting point for both the
construction of low energy effective theories and the computation of
macroscopic observables within the kinetic or hydrodynamical limits.

The global properties of a color superconductor depend dramatically
 on the number of quark flavors $N_f$ participating in the
condensation.  Here we shall discuss the case of two flavors $N_f=2$,
which exhibits a very rich structure. The diquark condensate breaks the
color $SU(3)$ group into an $SU(2)$ subgroup. As a result, five gluons
acquire masses through the Anderson-Higgs mechanism, while three of them
remain massless. Also  not all the quarks
attain a gap. 
In the high-density limit, and because of asymptotic freedom, the gap and
the gluon masses can be computed reliably from first principles
\cite{Son:1999,Schafer:1999b,Pisarski:2000bf,Hong:2000tn,Brown:2000aq,Manuel:2000nh,Rischke:2000qz}. A
well-defined hierarchy of scales then appears: at weak coupling $g\ll 1$,
the gap $\Delta$ is exponentially smaller than the gluon masses $\sim
g\mu$ and the chemical potential $\mu$.

The low energy physics of a two-flavor color superconductor is
dominated by its light degrees of freedom. At vanishing temperature,
these are the massless gauge bosons, the gapless quarks and a
(pseudo-) Goldstone boson, similar to the $\eta$ meson. However, the
gapless quarks and the $\eta$ meson are neutral with respect to the
unbroken $SU(2)$ subgroup. In turn, the condensate, although neutral
with respect to the unbroken $SU(2)$, polarizes the medium since their
constituents carry $SU(2)$ charges. Hence, the dynamics of the light
$SU(2)$ gauge fields differs from the vacuum theory. This picture has
recently been introduced by Rischke, Son and Stephanov
\cite{Rischke:2000cn}. Their infrared effective theory for momenta $k
\ll \Delta$ is
\begin{equation}\label{RSS}
S_{\rm eff}^{T=0} = \int d^4 x \left ( \frac{\epsilon}{2} \,
{\bf E}_a \cdot {\bf E}_a - \frac{1}{2\lambda} \,{\bf B}_a \cdot {\bf B}_a
\right) \ ,
\label{Seff-T0}
\end{equation}
where $E_i^a \equiv F_{0i}^a$ and $B_i^a \equiv \frac12 \epsilon_{ijk}
F_{jk}^a$ are the $SU(2)$ electric and magnetic fields. The constants
$\epsilon$ and $\lambda$ are the dielectric susceptibility and  magnetic
permeability of the medium. To leading order, $\lambda = 1$ and $\epsilon
= 1+ g^2 \mu^2/(18 \pi^2 \Delta^2)$ \cite{Rischke:2000cn}. As a
consequence,  the velocity of the $SU(2)$ gluons is smaller than in
vacuum. This theory is confining, but the scale of confinement is highly
reduced with respect to the one in vacuum with $\Lambda'_{\rm QCD} \sim
\Delta \exp{(-\frac{2 \sqrt{2} \pi}{11} \frac{\mu}{g \Delta})}$
\cite{Rischke:2000cn}. Due to asymptotic freedom, it is expected that
perturbative computations are reliable for energy scales larger than
$\Lambda'_{\rm QCD}$.   

At non-vanishing temperature, thermal excitations modify the low
energy physics.  The condensate melts at the critical
temperature $T_c \approx 0.567 \Delta_0$ \cite{Pisarski:2000bf}
($\Delta_0$ is the gap at vanishing temperature). We restrict the
discussion to temperatures within
$\Lambda'_{\rm QCD} \ll T <T_c$ , which provides the basis for the
perturbative computations below. In this regime, the main contribution
to the long distance properties of the $SU(2)$ fields stems from the
thermal excitations of the constituents of the diquark condensate. The
thermal excitations of the massless gauge fields contribute only at
the order $g^2T^2$ and are subleading for sufficiently large $\mu$.
Those of the gapless quarks and of the $\eta$ meson do not couple to
the $SU(2)$ gauge fields.

The thermal excitations due to the constituents of the diquark condensate
display a quasiparticle structure. This implies that they can be cast into
a transport equation. To that end, and working in natural units $k_B =
\hbar = c=1$, we introduce the on-shell one-particle phase space density
$f(x,{\bf p},Q)$, $x^\mu=(t,{\bf x})$, describing the quasiparticles. The
distribution function depends on time, the phase space variables position
${\bf x}$, momentum ${\bf p}$, and on $SU(2)$ color charges $Q_a$, with
the color index $a =1,2$ and $3$. The quasiparticles carry $SU(2)$ color
charges simply because the constituents of the condensate do. The on-shell
condition for massless quarks $m_q=0$ relates the energy of the
quasiparticle excitation to the chemical potential  and the gap as 
$p_0 \equiv \epsilon_p = \sqrt{(p-\mu)^2 + \Delta^2(T)}$. 
The gap is both temperature and momentum-dependent. From now on, we can
neglect its momentum dependence which is a subleading effect.  The
velocity of the quasiparticles is given by 
\beq\label{velocity}
{\bf v}_p 
\equiv \frac{ \partial \epsilon_p}{\partial {\bf p}} =
\frac{|p-\mu|}{\sqrt{(p-\mu)^2 + \Delta^2(T)}}\, {\hat {\bf p} \ ,} 
\eeq
and depends on both the chemical potential and the gap. For $\Delta=0$,
the quasiparticles would travel at the speed of light. However, in the
presence of the gap $\Delta\neq 0$, their propagation is suppressed, $v_p
\equiv |{\bf v}_p|\leq 1$. 

The one-particle distribution function $f(x,{\bf p},Q)$ obeys a very
simple transport equation, given by  
\begin{equation}\label{transport}
\left[D_t + {\bf v}_p \cdot {\bf D} -g Q_a \left({\bf E}^a +
{\bf v}_p \times {\bf B}^a \right) \frac{\partial}{\partial {\bf p}}
    \right] f = C[f] \,.
\end{equation}
Here, we  have introduced the short-hand notation $D_\mu f\equiv [\partial
_\mu-g \epsilon^{abc}Q_c A_b^\mu{\partial ^Q_a}]f$ for the covariant
derivative acting on $f$. The first two terms on the l.h.s.~of
\Eq{transport} combine to a covariant drift term $v_p^\mu D_\mu$, where
$v_p^\mu = (1, {\bf v}_p)$ and $D_\mu =(D_t, {\bf D})$. The terms
proportional to the color electric and magnetic fields provide a force
term. The r.h.s.~of \Eq{transport} contains a (yet unspecified) collision
term $C[f]$. 

The thermal quasiparticles carry an $SU(2)$ charge, and hence provide an
$SU(2)$ color current. It is given by
\begin{equation}\label{current}
J^\mu_a(x)=g \sum_{\hbox{\tiny helicities}\atop\hbox{\tiny species}}
\int \frac{d^3 p}{(2\pi)^3}
dQ \ v^\mu_p\, Q_a\, f(x,{\bf p},Q) \ . 
\end{equation}
Below, we simply omit a species or helicity index on $f$, as well as the
explicit sum over them. The color measure is normalised $\int dQ=1$, and
obeys $\int dQQ_a=0$ and $\int dQQ_aQ_b=C_2\delta_{ab}$, where $C_2$
denotes the quadratic Casimir ($C_2=\s012$ for quarks in the
fundamental). The color current \Eq{current} is covariantly conserved for
$C[f]=0$. For $C[f] \neq 0$
 a covariantly conserved current implies
certain restrictions in the form of the collision term.

Transport equations similar to \Eq{transport} have been known
previously: (i) The analogue of \Eq{transport} at high temperature for
$\mu=\Delta=0$ (and $SU(N)$ color charges $Q_a$) has been introduced in
\cite{Heinz} for the description of the quark-gluon plasma at high
temperature. In the Vlasov approximation and for $C[f]=0$, it reproduces
the hard thermal loop (HTL) effective theory 
\cite{HTL}
 to leading order in $g$ \cite{KLLM}. (ii) A formalism which allows to go
systematically beyond the Vlasov approximation has been derived as well
\cite{LM}. This includes the derivation of the collision term $C[f]$. 
(iii) At high density and $T=\Delta=0$, the corresponding transport
equation leads to the  hard dense loop (HDL) effective theory for
$C[f]=0$, and to leading order in $g$ \cite{Manuel:1996td}. (iv) In the
non-relativistic limit for the Abelian case, our transport equation
reduces to the one for a BCS superconductor \cite{K}.

In the remaining part of the Letter, we study the collisionless dynamics
$C[f] =0$ of the  color superconductor close to thermal equilibrium and to
leading order in the gauge coupling. Consider the distribution function
$f(x,{\bf p},Q)=  f^{\rm eq}(p_0)+ g  f^{(1)}(x,{\bf p},Q)$. Here, $f^{\rm
eq.}(p_0) =
1/ (\exp(\epsilon_p/T) +1)$ is the fermionic equilibrium distribution
function and $g  f^{(1)}(x,{\bf p},Q)$  describes a slight deviation from
equilibrium. For convenience, we also introduce the color density
\begin{equation}\label{currentdensity}
J_{a}(x,{\bf p})= g  \!\int\! dQ Q_{a} f(x,{\bf p},Q) \ ,
\end{equation}
from which the induced color current of the medium \Eq{current} follows as
$J^\rho _a(x) = \int \s0{d^3p}{(2\pi)^3}v_p^\rho J_{a}(x,{\bf
p})$. Expanding the transport equation \eq{transport} to leading order in
$g$, and taking the two helicities per quasiparticle into account, we find
the transport equation for the color density as
\begin{equation}\label{transport-current}
\left[D_t + {\bf v}_p \cdot {\bf D}\right]   J(x,{\bf p})  =
g^2 N_f  \,  {\bf v}_p \cdot  {\bf E}(x)\, \frac{d f^{\rm eq}}{d
\epsilon_p} \ .
\end{equation}
The solution of the transport equation reads
\begin{eqnarray}
\label{current-solution}
J^\mu_a(x) =
g^2 N_f 
\!\int\! 
\frac{d^3 p\, d^4y}{(2\pi)^3}\,
v_p^\mu \, G_{ab} \,
{\bf v}_p \cdot {\bf E}_b (y)
\frac{d f^{\rm eq}}{d \epsilon_p} 
.
\end{eqnarray}
with the Greens function 
$G_{ab}\equiv \langle y|1/(v_p \cdot D)|x\rangle_{ab}$.
After having solved the transport equation, the relevant information
concerning the low energy effective theory is contained in  the functional
$J[A]$. Notice that the above derivation is analogous to the derivation of
the HTL and HDL effective theories from kinetic theory
\cite{HTL,KLLM,Manuel:1996td}.
 Owing to this resemblance, we call the diagrams which are derived from
\Eq{current-solution} as {hard superconducting loops} (HSL). The HSL
effective action follows from \Eq{current-solution} by solving $J[A] =
-{\delta \Gamma_{\rm HSL}[A]}/{\delta A}$ for $\Gamma_{\rm HSL}[A]$, and
all HSL diagrams can be derived by performing functional derivatives to
the effective action (or the induced current). We thus reach to the
conclusion that the low energy effective theory for a two-flavor color
superconductor at finite temperature reads $S_{\rm eff}^T = S_{\rm
eff}^{T=0} + \Gamma_{\rm HSL}$ to leading order in $g$. This theory is
effective for modes with $k\ll \Delta$.

Let us have a closer look into the induced current, which we formally
expand as
$J^a_\mu[A]
=
 \Pi^{ab}_{\mu\nu} A^\nu_b
+\s012\Gamma^{abc}_{\mu\nu\rho} A^\nu_b  A^\rho_c
+\ldots$ 
in powers of the gauge fields. The most relevant information on the
thermal effects is contained in the thermal polarization tensor
$\Pi^{ab}_{\mu\nu}$. Using \Eq{current-solution}, we find
$$
\Pi^{\mu \nu}_{ab}(k) 
= 
g^2 N_f \delta_{ab}  
\int \frac{d^3 p}{(2 \pi)^3} 
\frac{d f^{\rm eq}}{d \epsilon_p}
\left( g^{\mu 0} g^{\nu 0} - k_0  \frac{v_p^\mu v_p^\nu}{k \cdot
v_p}\right) 
\,.
$$
It obeys the Ward identity $k_\mu \Pi^{\mu \nu}_{ab}(k) =0$. With retarded
boundary conditions $k_0 \rightarrow k_0 + i 0^+$, the  polarization
tensor has an imaginary part, 
$$
{\rm Im}\, \Pi^{\mu \nu}_{ab} (k) 
= \delta_{ab}  \pi g^2 N_f  k_0 
   \int \!\!\ \frac{d^3 p}{(2 \pi)^3} 
\frac{d f^{\rm eq}}{d \epsilon_p}
    v_p^\mu v_p^\nu \,\delta (k \cdot v_p) \,,            
$$
which corresponds to  Landau damping. Performing the angular integration, 
we obtain for the  longitudinal and transverse
projections of the polarization tensor
\begin{mathletters}\label{pipi}
\begin{eqnarray}
\Pi_{L} (k) & = & 
\frac{g^2 N_f}{2 \pi^2} 
\int^\infty _0 \!\!\! dp\, p^2 
\frac{d f^{\rm eq}}{d \epsilon_p} 
\!\! \left[ 1 - \frac{1}{2}\frac{k_0}{k v_p}\times \right.
\nonumber
\\
\label{Pi-L}
&\times & \left. 
\left(\,{\rm ln\,}\left| \frac{k_0+ k v_p}{k_0- k v_p} \right| 
-i \pi \, \Theta( k^2 v_p^2 -k_0^2) \right)  \right] \ , 
\\
\Pi_{T} (k) & = & 
\frac{g^2 N_f}{4 \pi^2}\frac{k_0^2}{k^2}  
\int^\infty _0 \!\!\! dp\, p^2 
\frac{d f^{\rm eq.}}{d \epsilon_p}
   \!\!  \left[ 1 + \frac12 \left( \frac{k v_p}{k_0} -
\frac{k_0}{k v_p} \right) \times\right. 
\nonumber \\
\label{Pi-T}
& \times& \left. \left( {\rm ln\,} \left|{\frac{k_0+
k v_p}{k_0- k v_p}}\right|
 -  i \pi \, \Theta( k^2 v_p^2 -k_0^2)
 \right) \, \right] \ ,
\end{eqnarray}
\end{mathletters}%
where $\Theta$ is the step function.
We first consider the real part of the polarization tensor. From
\Eq{pipi}, and in the limit $k_0\to 0$,  we infer  that the  longitudinal
gauge bosons acquire a thermal mass, the Debye mass, while the transverse
ones remain massless. The (square of the) Debye mass is given by
\begin{equation}\label{Debye}
m^2_D = -\frac{g^2 N_f}{2 \pi^2} \int^\infty _0 \!\!\! dp\,p^2 
\frac{d f^{\rm eq.}}{d \epsilon_p} \equiv M^2
\, I_0\left(\0{\Delta}{T},\0{T}{\mu}\right)\,. 
\end{equation}
For convenience, we have factored-out the Debye mass $M$ of the
ultradegenerate plasma in the normal phase, $M^2 \equiv {g^2 N_f \mu^2}/{2
\pi^2}$. The dimensionless functions 
\begin{equation}
I_n \left(\0{\Delta}{T},\0{T}{\mu}\right)=
-\frac{1}{\mu^2}\int^\infty _0 \!\!\! dp\,
{p^2}\frac{d f^{\rm eq.}}{d \epsilon_p} 
v_p^n 
\end{equation}
obey $I_n\ge I_{n+1} >0$ for all $n$ due to $v_p \leq 1$. Equality holds
for vanishing gap. For the physically relevant range of parameters $T <
\Delta \ll \mu$, the functions  $I_n$ are $ \ll 1$. In particular, it is
easy to see that $I_n(\infty,0)=0$: there is no Debye screening for the
$SU(2)$ gluons at $T=0$ in the superconducting phase. In the limit where
$\Delta/T \gg 1$, and to leading order in $T/\mu\ll 1$, the Debye mass
reduces to
\begin{equation}\label{Debye2}
m^2_D = M^2 \, 
\sqrt{2 \pi \frac{\Delta}{T}}
\exp(-{\Delta}/{T}) \,. 
\end{equation} 
The dispersion relations for the longitudinal and transverse gluons follow
from the poles of the corresponding propagators,
\begin{mathletters}\label{DR}
\begin{eqnarray}
\epsilon k^2 
- {\rm Re}\, \Pi_L(k_0,k) \Large|_{k_0=\omega_L(k)} 
& = & 0 \ ,
\\
\epsilon k^2_0
- \01{\lambda}k^2 
+ {\rm Re}\,\Pi_T(k_0,k)\Large|_{k_0=\omega_T(k)}
& = & 0 \ ,
\end{eqnarray}
\end{mathletters}%
where $\epsilon$ and $\lambda$ have been introduced in \Eq{RSS}. The
plasma frequency $\omega_{\rm pl}$ follows from \Eq{DR} as
\beq\label{PlasmaFrequency}
\omega^2_{\rm pl}=
\01{3\epsilon} M^2 \, I_2\left(\0{\Delta}{T},\0{T}{\mu}\right)\,.
\eeq
For generic external momenta the dispersion relations can only be solved
numerically. In turn, if the spatial momenta are much smaller than the
plasma frequency $k \ll \omega_{\rm pl}$, solutions to \Eq{DR} can be
expanded in powers of $k^2/\omega^2_{\rm pl}$ as
\begin{mathletters}\label{DR-Expansion}
\begin{eqnarray}
\omega^2_L(k) 
& = &
\omega^2_{\rm pl}
\left[
1 
+ \035\frac{I_4}{I_2} \frac{k^2}{\omega^2_{\rm pl}} 
+ {\cal O} (\frac{k^4}{\omega^4_{\rm pl}}) 
\right]\ , \\
\omega^2_T(k) 
& = &
\omega^2_{\rm pl} 
\left[
1
+ \left(\01{\epsilon\lambda}+ \015\frac{I_4}{I_2} \right)
  \frac{k^2}{\omega^2_{\rm pl}} 
+ {\cal O} (\frac{k^4}{\omega^4_{\rm pl}})  \right]\,.
\end{eqnarray}
\end{mathletters}%
Let us now consider the imaginary part of \Eq{pipi}, which describes
Landau damping. Since $v_p \leq 1$, we conclude that Landau damping only
occurs for $k^2_0 \leq k^2$. Hence, plasmon and transverse gluon
excitations are  stable as long as $\omega_{L,T}(k)>k$. Furthermore, we
notice that the imaginary part of \Eq{pipi} is logarithmically
divergent: the quasiparticle velocity vanishes for momenta close to the
Fermi surface, which is an immediate consequence of the presence of a gap,
cf.~\Eq{velocity}. This divergence does not appear in the real part,
because the logarithm acts as a regulator for the $1/v_p$ factor. To
leading order in $T/\mu$, and in the region of small frequencies $k_0^2
\ll k^2$, we find at logarithmic accuracy, and for all values of
$\Delta/T$,
\begin{mathletters}\label{ImPi}
\begin{eqnarray}
{\rm Im}\, \Pi_L &=& 
- 2\pi M^2 \frac{k_0}{k}\0{\Delta}{T}
\frac{  \ln{(k/k_0)}}{(e^{\Delta/T} +1)(e^{-\Delta/T} +1)} \ ,
\\ 
{\rm Im}\, \Pi_T &=&  
\pi M^2 
\frac{k_0}{k}
\left[
\frac{1}{e^{\Delta/T} +1} 
\right.  
\nonumber\\
& - & 
\left. 
2\frac{k^2_0}{k^2}  \0{\Delta}{T}
 \frac{ \ln{k/k_0}}{(e^{\Delta/T} +1)(e^{-\Delta/T} +1)} 
\right] \ .
\end{eqnarray}
\end{mathletters}%
For small frequencies, Landau damping is dominated by the logarithmic
terms, which are proportional to the gap.  Once the gap vanishes,
subleading terms in $\s0{\Delta}{T}$, not displayed in \Eq{ImPi}, take
over and reduce ${\rm Im}\Pi$ to known expressions for the normal
phase.

Finally, we explain how the polarization tensor, as obtained within
the present transport theory, matches the computation of
$\Pi^{\mu\nu}$ for external momenta $k_0, k \ll \Delta$ to one-loop
order from quantum field theory. The one-loop gluon self-energy for a
two-flavor color superconductor has been computed by Rischke, and the
polarization tensor for the unbroken $SU(2)$ subgroup is given in
Eq.~(99) of \cite{Rischke:2000qz}. It contains contributions from
particle-particle, particle-antiparticle and antiparticle-antiparticle
excitations.  The particle-antiparticle contribution to $\Pi^{00}$ and
$\Pi^{0i}$ at low external momenta, and the antiparticle-antiparticle
excitations are subleading. The particle-particle contributions divide
into two types. The first ones have poles for gluonic frequencies $k_0
= \pm \left(\epsilon_p + \epsilon_{p-k} \right)$ and an imaginary part
once $k_0$ exceeds the Cooper pair binding energy $2 \Delta$.  These
terms are related to the formation or breaking of a Cooper pair, and
suppressed for low external gluon momenta.  The second type of terms,
only non-vanishing for $T \neq 0$, have poles at $k_0 = \pm
\left(\epsilon_p - \epsilon_{p-k} \right)$. For $k \ll \Delta$ we
approximate it by $k_0 \approx \pm \frac{\partial \epsilon_p}{\partial
  {\bf p}} \cdot {\bf k}$. The prefactor, a difference of thermal
distribution functions, is approximated by $f^{\rm eq}(\epsilon_p) -
f^{\rm eq}(\epsilon_{p-k})\approx \frac{\partial \epsilon_p}{\partial
  {\bf p}} \cdot {\bf k} \,\frac{d f^{\rm eq}}{d \epsilon_p}$. After
simple algebraic manipulations we finally end up with the result
given above. We conclude that this part of the one-loop
polarization tensor describes the collisionless dynamics of thermal
quasiparticles for a two-flavor color superconductor.  The same type
of approximations can be carried out for $\Pi^{ij}$ to one-loop order.
There, apart from the HSL contributions, additional terms arise due to
particle-particle and particle-antiparticle excitations, cf.~Eq.~(112)
of \cite{Rischke:2000qz}. We have not evaluated these terms
explicitly. However, we expect them to be subleading or vanishing, as
otherwise the Ward identity $k_\mu \Pi^{\mu \nu}_{ab}(k) =0$ is
violated. For $T=0$, this has been confirmed in \cite{Rischke:2000qz}.

Summarizing, we have provided a transport equation for the gapped
quarks of two-flavor color superconductors. Its simple structure is
based on the quasiparticle behavior of the thermal excitations of the
condensate, in consistency with the underlying quantum field theory.
We have constructed a low temperature infrared effective theory of the
superconductor.  To leading order, we found Landau damping, and Debye
screening of the chromo-electric fields. Beyond leading order,
chromo-magnetic fields are damped because they scatter with the
quasiparticles. The damping rate is related to the color conductivity.
It should be possible to compute the rate from the transport
equation~(\ref{transport}), amended by the relevant collision term.
The latter can be derived, for example, using the methods developped
in \cite{LM}.

We have not discussed the transport equations for gapless quarks or
the $\eta$ meson, because they do not carry $SU(2)$ charges.  However,
their excitations are light compared to the gapped quasiparticles, and
dominant for other transport properties such as thermal and electrical
conductivities or shear viscosity. The corresponding set of transport
equations will be discussed elsewhere.

It would be very interesting to study the transport equations in a
three-flavor color superconductor \cite{Alford:1999mk}.  For $N_f=3$
the quark-quark condensate breaks the $SU(3)$ gauge group completely,
as well as some global flavor symmetries.  Transport phenomena should
then be dominated by the Goldstone modes associated to the breaking of
the global symmetries.  The corresponding transport equations will be
substantially different for the two and three flavor case.
\\[-.5ex]

This work has been supported by the European Community through the
Marie-Curie fellowships HPMF-CT-1999-00404 and HPMF-CT-1999-00391.

\end{document}